%% file: main.tex
\documentclass[12pt]{article}
\usepackage[utf8]{inputenc}
\usepackage{authblk}
\usepackage{setspace}
\usepackage[margin=1.25in]{geometry}
\usepackage{graphicx}
\usepackage{subcaption}
\usepackage{amsmath}
\usepackage{lineno}
\usepackage{lipsum}
\usepackage[style=nejm, 
citestyle=numeric-comp,
sorting=none]{biblatex}
\input{command}
\begin{document}
\input{content}
\end{document}

%% file: content.tex
\title{Inclusive photon multiplicity at forward pseudorapidities in pp and p--Pb collisions \\ at $\sqrt{s_{\rm NN}}$ = 5.02~TeV with ALICE}
\author[1*]{Abhi Modak (for the ALICE Collaboration)}
\affil[1]{Bose Institute, Kolkata, India}
\affil[*]{Address correspondence to: abhi.modak@cern.ch}

\onehalfspacing
\maketitle

\date{}

\begin{abstract}

Global observables such as the pseudorapidity distributions of particle multiplicities in the final state are crucial to shed light into the physics processes involved in hadronic collisions. In proton--lead (p--Pb) collisions at Large Hadron Collider (LHC) energies, such measurements provide an important baseline to understand lead--lead (Pb--Pb) results by disentangling hot nuclear matter effects from the ones due to the cold nuclear matter. Multiplicity measurements can also put constraints on theoretical models describing the initial stages of the collision, e.g., to what degree the nucleon and the nuclei interact as dilute (partons) or dense (CGC-like) fields. The study of inclusive photon multiplicity aims to provide complementary measurements to those obtained with charged particles.

In these proceedings, the pseudorapidity distributions of inclusive photons at forward pseudorapidity (2.3~$<~\eta_{\rm \,lab}~<$~3.9) in pp and p--Pb collisions at $\sqrt{s_{\rm NN}}$ = 5.02 TeV are presented. The data samples were collected using the Photon Multiplicity Detector (PMD) of ALICE. The multiplicity dependence of photon production in p--Pb collisions is presented and a comparison with charged-particle distributions measured at mid-pseudorapidity is shown. The results are also compared with predictions from Monte Carlo event generators.

\end{abstract}

\section{Introduction}
One of the primary goals of heavy-ion collision experiments, such as ALICE, is to study and understand the properties of the deconfined state of nuclear matter, commonly known as the quark--gluon plasma (QGP). The first step in characterizing the produced QGP matter in these collisions is the measurement of pseudorapidity distributions of produced final-state particles. Such studies in pp and p--Pb collisions are also important as they provide the baselines for the interpretation of the measurements in heavy-ion collisions. This contribution reports the measurements of inclusive photon multiplicities for minimum bias pp, p--Pb collisions and for various multiplicity classes in p--Pb collisions at $\sqrt{s\rm_{NN}}$~=~5.02~TeV.

\section{Data analysis}

This analysis was performed using the ALICE~\cite{ALICE:Exp} data collected in 2013 during LHC Run~1 for p--Pb collisions and in 2015 during LHC Run~2 for pp collisions. The data from p--Pb collisions were recorded for two beam configurations: in one (denoted as p--Pb), the lead beam travelled towards positive $\eta_{\rm \,lab}$ and in the other configuration (denoted as Pb--p) it moved towards negative $\eta_{\rm \,lab}$. The pp data were analysed for inelastic events, whereas measurements in p--Pb collisions were performed for non-single diffractive interactions. Events with the reconstructed primary vertex position along the beam line, $|v_{z}|<10$ cm, from the nominal interaction point were considered. The multiplicity classes were determined by measuring the charged-particle multiplicity in the outer layer of the Silicon Pixel Detector~\cite{spd} at mid-pseudorapidity (denoted as CL1 estimator) and the energy deposited in the Pb-remnant side of the neutron calorimeter~\cite{zdc} at beam rapidity (denoted as ZNA estimator)~\cite{ALICE:ChPrpPbCent}. The raw distributions of photons were obtained by counting the number of reconstructed clusters (in the preshower plane of the Photon Multiplicity Detector (PMD)~\cite{pmd}) that satisfied the photon--hadron discrimination thresholds~\cite{ALICE:PMDpp,ALICE:PMDpPb}. The distributions were then corrected for various instrumental effects (detector inefficiency, limited acceptance, contaminations from hadron clusters and secondary particles produced in interactions with surrounding materials of the PMD) using a Bayesian unfolding method~\cite{BayesUnfold}. Systematic uncertainties from various sources (effect of upstream material in front of the PMD, hadron and secondary photon contaminations, event generator dependence, unfolding method) were estimated and then added in quadrature. The total systematic uncertainty was found to be around 9--10\%~\cite{ALICE:PMDpPb}.

\begin{figure}
\begin{subfigure}[b]{0.4\textwidth}
   \includegraphics[scale=0.4]{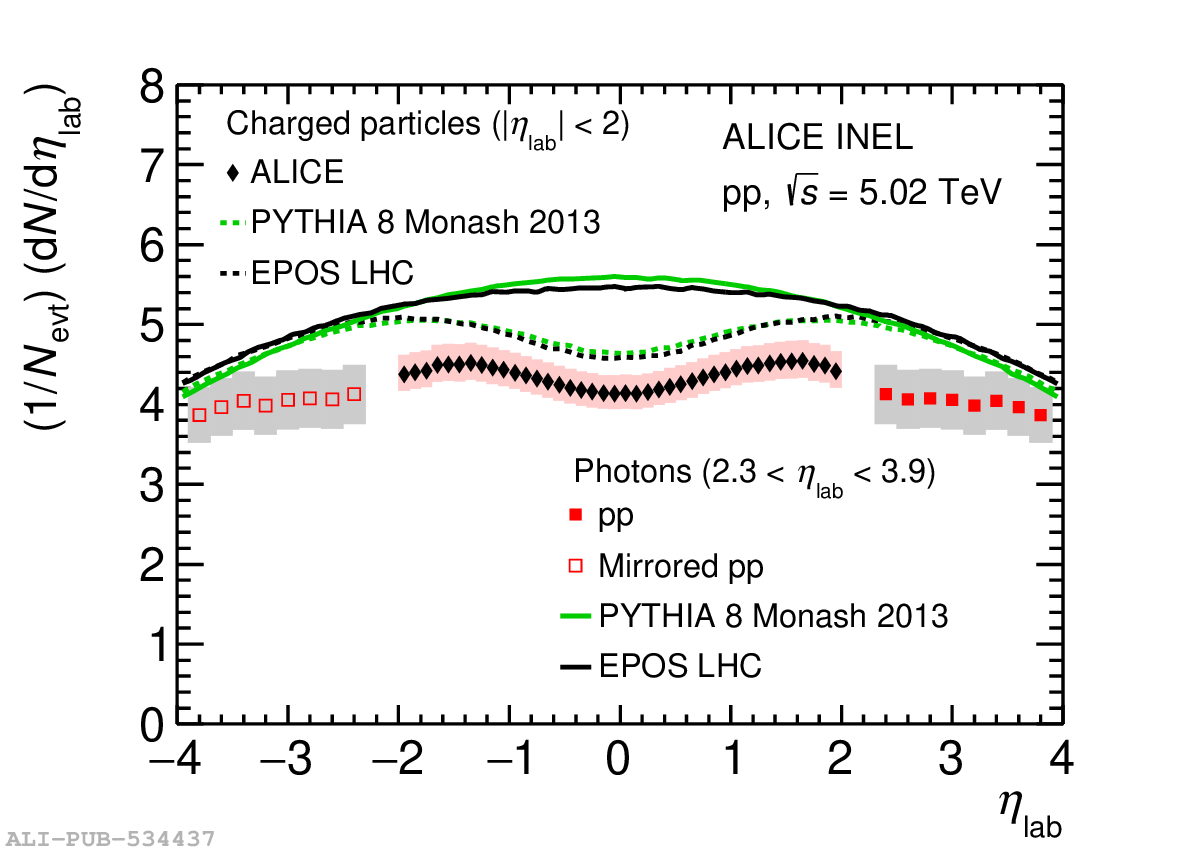}
    \caption{}
    \label{dndeta_pp}
  \end{subfigure}
  \hspace*{1.2cm}
   \begin{subfigure}[b]{0.4\textwidth}
    \includegraphics[scale=0.4]{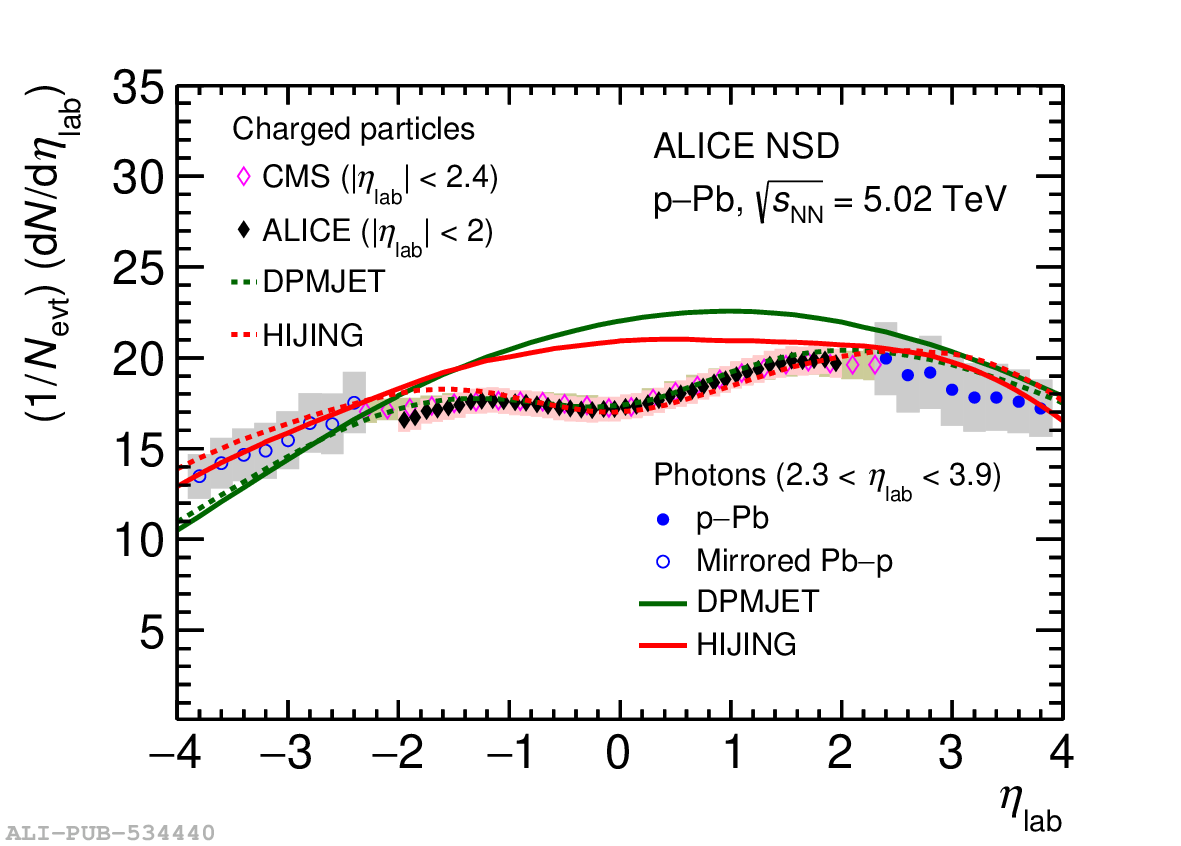}
    \caption{}
    \label{dndeta_pPb}
  \end{subfigure}
  \caption{The d$N_{\rm \gamma}$/d$\eta\rm_{lab}$ measured in pp (solid red) (a) and p--Pb (solid blue) (b) collisions at $\sqrt{s\rm_{NN}}$~=~5.02~TeV. The pp and Pb--p data points have been reflected around $\eta_{\rm \,lab}$ = 0 and shown by open red and open blue markers, respectively. The results are compared with similar distributions of charged particles at mid-pseudorapidity~\cite{ALICE:ChPrppMB,ALICE:ChPrpPbMB,CMS:ChPrpPbMB} and with various MC predictions. Shaded regions show systematic uncertainties.}
  \label{dndeta_MB}
\end{figure}

\section{Results and discussion}

Figure~\ref{dndeta_MB} presents the pseudorapidity distributions (d$N_{\rm \gamma}$/d$\eta_{\rm \,lab}$) of inclusive photons in pp, p--Pb, and Pb--p collisions at $\sqrt{s_{\rm NN}}$~=~5.02~TeV measured within 2.3~$<~\eta_{\rm \,lab}~<$~3.9 together with the measurements of charged-particle multiplicities (d$N_{\rm ch}$/d$\eta_{\rm \,lab}$) at mid-pseudorapidity~\cite{ALICE:ChPrppMB,ALICE:ChPrpPbMB,CMS:ChPrpPbMB}. The data from pp and Pb--p collisions are reflected around $\eta_{\rm \,lab}$~=~0 to extend the measurements in the region, $-3.9<\eta_{\rm \,lab}<-2.3$. The d$N_{\rm \gamma}$/d$\eta_{\rm \,lab}$ at forward pseudorapidity smoothly matches with the d$N_{\rm ch}$/d$\eta_{\rm \,lab}$ at mid-pseudorapidity indicating that the production mechanisms for charged and neutral pions are similar. The predictions from various MC models are also displayed in Fig.~\ref{dndeta_MB} and show similar values for photon (solid lines) and charged-particle (dashed lines) multiplicities at forward and backward pseudorapidities, while at mid-pseudorapidity the d$N_{\rm \gamma}$/d$\eta_{\rm \,lab}$ differs from the d$N_{\rm ch}$/d$\eta_{\rm \,lab}$. This difference is due to a mass effect in the transformation between $\mathrm{d}N/\mathrm{d}y$ and $\mathrm{d}N/\mathrm{d}\eta$ at $\eta \approx 0$. Both HIJING (v1.36)~\cite{hijing} and DPMJET (v3.0-5)~\cite{dpmjet} event generators fairly describe the measured d$N_{\rm ch}$/d$\eta_{\rm \,lab}$ in p--Pb collisions. The DPMJET slightly underpredicts the d$N_{\rm \gamma}$/d$\eta_{\rm \,lab}$ in the p-going side and reproduces the same within uncertainties in the Pb-going side. For pp collisions, both EPOS LHC~\cite{eposlhc} and PYTHIA 8 (v8.243) with the Monash 2013~tune~\cite{pythia8_monash} overestimate the photon and charged-particle multiplicity.

\begin{figure}[h!]
  \begin{minipage}[c]{0.5\textwidth}
    \includegraphics[scale=0.45]{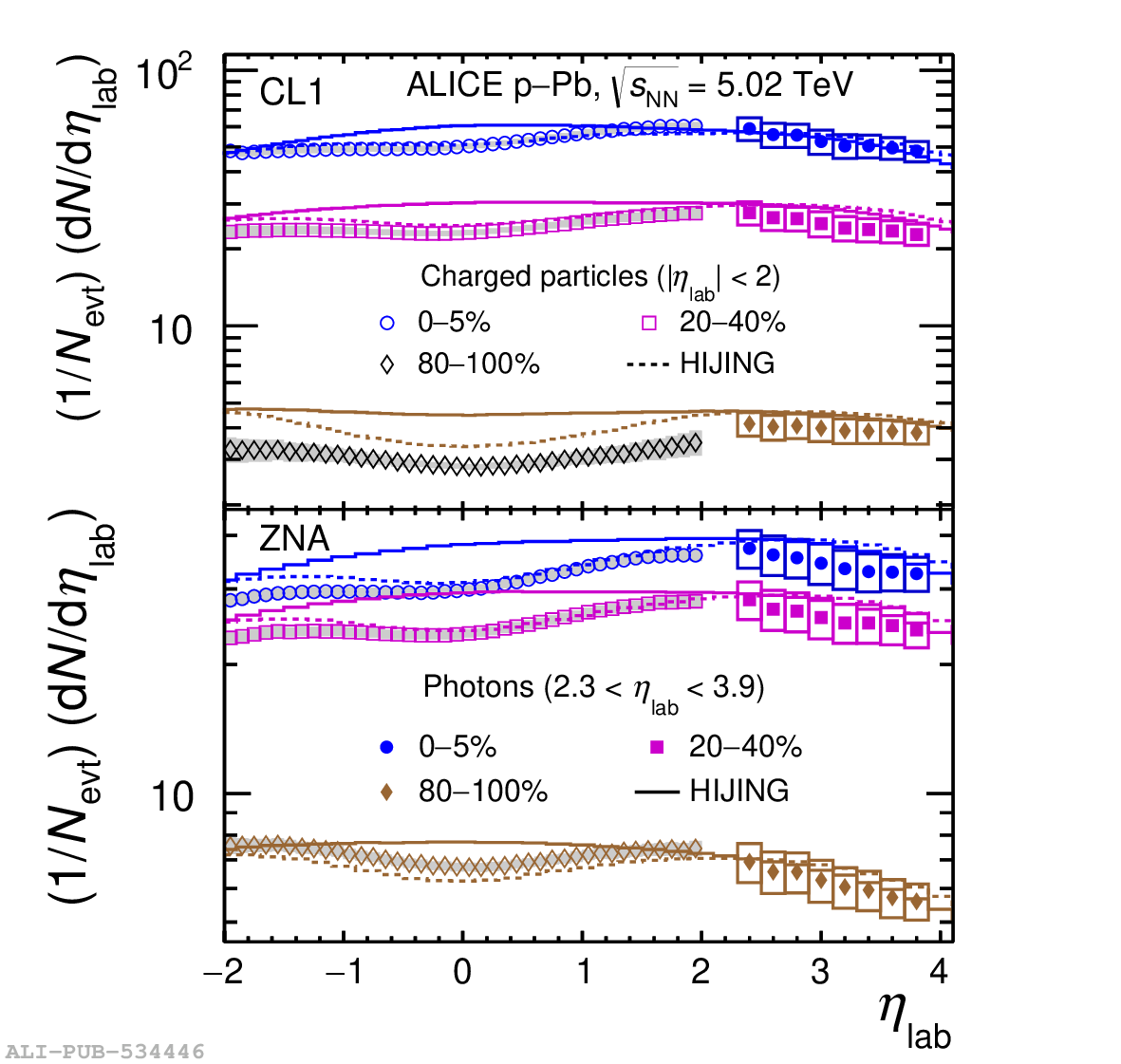}
  \end{minipage}\hfill
  \begin{minipage}[c]{0.45\textwidth}
   \caption{The d$N_{\rm \gamma}$/d$\eta_{\rm \,lab}$ (solid markers) measured within 2.3~$<~\eta_{\rm \,lab}~<$~3.9 for different multiplicity classes using two multiplicity estimators: CL1 (top), ZNA (bottom) in p--Pb collisions at $\sqrt{s_{\rm NN}}$~=~5.02~TeV and a comparison with similar studies of charged-particle multiplicities (open markers) at mid-pseudorapidity~\cite{ALICE:ChPrpPbCent}. Predictions from the HIJING event generator are superimposed. Shaded regions and open boxes show the systematic uncertainties for charged particles and photons, respectively.}
   \label{dndeta_cent}
   \end{minipage}
\end{figure}

Figure~\ref{dndeta_cent} shows the pseudorapidity distributions of both photons and charged particles measured in p--Pb collisions for three multiplicity classes (0--5\%, 20--40\% and 80--100\%) determined with the CL1 (top panel) and ZNA (bottom panel) estimators. The particle density in the highest multiplicity class (0--5\%) when considering the CL1 (ZNA) estimator reaches values thrice (twice) as large as those in minimum bias p--Pb collisions. A clear asymmetric shape is observed for d$N_{\rm ch}$/d$\eta_{\rm \,lab}$ in the highest multiplicity class (0--5\%) and the shape becomes symmetric, like in pp, in the lowest multiplicity class (80--100\%). HIJING describes the d$N_{\rm \gamma}$/d$\eta_{\rm \,lab}$ at forward pseudorapidity within the measurement uncertainties. For 80--100\% event class, HIJING overestimates (underestimates) the d$N_{\rm ch}$/d$\eta_{\rm \,lab}$ for the CL1 (ZNA) estimator.

\section{Conclusion}

The pseudorapidity distributions of inclusive photons were measured over a kinematic region of 2.3~$<~\eta_{\rm \,lab}~<$~3.9 for minimum bias pp, p--Pb, and Pb--p collisions and for different multiplicity classes in p--Pb collisions at $\sqrt{s_{\rm NN}}$~=~5.02~TeV. The d$N_{\rm \gamma}$/d$\eta_{\rm \,lab}$ at forward pseudorapidity was observed to follow the trend of similar measurements of charged particles at mid-pseudorapidity. The predictions from various MC describe the data within 15--20\%. These results will help to establish baselines for the interpretation of Pb--Pb collision data.

\printbibliography

%% file: bibitems.bib
@article{ALICE:PMDpp,
        author = "Abelev, Betty Bezverkhny and others",
	authortype = {(ALICE Collaboration),},
	%title = "{Inclusive photon production at forward rapidities in proton--proton collisions at $\sqrt{s}~=$~0.9, 2.76 and 7~TeV}",
	eprint = "1411.4981",
	archivePrefix = "arXiv",
	primaryClass = "nucl-ex",
	doi = "10.1140/epjc/s10052-015-3356-2",
	journal = "Eur. Phys. J. C",
	volume = "75",
	number = "4",
	pages = "146",
	year = "2015"
}

@article{ALICE:PMDpPb,
    author = "Acharya, S. and others",
    authortype = {(ALICE Collaboration),},
    %title = "{ALICE Collaboration, Inclusive photon production at forward rapidities in pp and p$-$Pb collisions at $\sqrt{s_{\rm NN}} = 5.02$ TeV}",
    doi = "10.48550/arXiv.2303.00590",
    journal = "arXiv:2303.00590 [nucl-ex]",
    reportNumber = "CERN-EP-2023-026",
    month = "3",
    year = "2023"
}

@article{ALICE:ChPrpPbCent,
        authortype = {(ALICE Collaboration),},
        author = "Adam, Jaroslav and others",
	%title = "{Centrality dependence of particle production in p--Pb collisions at $\sqrt{s_{\rm NN}} = 5.02$ TeV}",
	eprint = "1412.6828",
	archivePrefix = "arXiv",
	primaryClass = "nucl-ex",
	doi = "10.1103/PhysRevC.91.064905",
	journal = "Phys. Rev. C",
	volume = "91",
	number = "6",
	pages = "064905",
	year = "2015"
}

@article{BayesUnfold,
    author = "G. D'Agostini",
    %title = "{A multidimensional unfolding method based on Bayes' theorem}",
    journal = "Nucl. Instrum. Methods A",
    volume = {362},
    number = {2},
    pages = {487-498},
    year = {1995},
    issn = {0168-9002},
    doi = "10.1016/0168-9002(95)00274-X"
}

@article{ALICE:ChPrppMB,
    author = "Acharya, S. and others",
    authortype = {(ALICE Collaboration),},
    %title = "{ALICE Collaboration, Pseudorapidity densities of charged particles with transverse momentum thresholds in pp collisions at $\sqrt{s} = 5.02$ and $13$ TeV}",
    doi = "10.48550/arXiv.2211.15364",
    journal = "arXiv:2211.15364 [nucl-ex]",
    reportNumber = "CERN-EP-2022-262",
    month = "11",
    year = "2022"
}

@article{ALICE:ChPrpPbMB,
	author = "Abelev, Betty Bezverkhny and others",
	authortype = {(ALICE Collaboration),},
	%title = "{Pseudorapidity density of charged particles in p--Pb collisions at $\sqrt{s_{\rm NN}} = 5.02$ TeV}",
	eprint = "1210.3615",
	archivePrefix = "arXiv",
	primaryClass = "nucl-ex",
	doi = "10.1103/PhysRevLett.110.032301",
	journal = "Phys. Rev. Lett.",
	volume = "110",
	number = "3",
	pages = "032301",
	year = "2013"
}

@article{CMS:ChPrpPbMB,
      author = "Sirunyan, Albert M and others",
      authortype = {(CMS Collaboration),},
      %title = "{Pseudorapidity distributions of charged hadrons in p--Pb collisions at $\sqrt{s_{\mathrm{NN}}}$ = 5.02 and 8.16~TeV}",
      eprint = "1710.09355",
      archivePrefix = "arXiv",
      primaryClass = "hep-ex",
      doi = "10.1007/JHEP01(2018)045",
      journal = "JHEP",
      pages = "045",
      volume = "01",
      year = "2018"
}

@article{hijing,
    author = "Wang, Xin-Nian and Gyulassy, Miklos",
    %title = "{HIJING: A Monte Carlo model for multiple jet production in pp, p--A and A--A collisions}",
    doi = "10.1103/PhysRevD.44.3501",
    journal = "Phys. Rev. D",
    volume = "44",
    pages = "3501--3516",
    year = "1991"
}

@article{dpmjet,
    author = "Roesler, Stefan and Engel, Ralph and Ranft, Johannes",
    %title = "{The Monte Carlo event generator DPMJET-III}",
    %booktitle = "{International Conference on Advanced Monte Carlo for Radiation Physics, Particle Transport Simulation and Applications (MC 2000)}",
    journal = "arXiv:hep-ph/0012252",
    doi = "10.48550/arXiv.hep-ph/0012252",
    month = "12",
    year = "2000"
}

@article{eposlhc,
    author = "Pierog, T and others",
    %title = "{EPOS LHC: Test of collective hadronization with data measured at the CERN Large Hadron Collider}",
    eprint = "1306.0121",
    archivePrefix = "arXiv",
    primaryClass = "hep-ph",
    doi = "10.1103/PhysRevC.92.034906",
    journal = "Phys. Rev. C",
    volume = "92",
    number = "3",
    pages = "034906",
    year = "2015"
}

@article{pythia8_monash,
    author = "Skands, Peter and Carrazza, Stefano and Rojo, Juan",
    %title = "{Tuning PYTHIA 8.1: the Monash 2013 Tune}",
    eprint = "1404.5630",
    archivePrefix = "arXiv",
    primaryClass = "hep-ph",
    doi = "10.1140/epjc/s10052-014-3024-y",
    journal = "Eur. Phys. J. C",
    volume = "74",
    number = "8",
    pages = "3024",
    year = "2014"
}

@article{ALICE:Exp,
	author = "Aamodt, K and others",
	authortype = {(ALICE Collaboration),},
	%title = "{The ALICE experiment at the CERN LHC}",
	doi = "10.1088/1748-0221/3/08/S08002",
	journal = "JINST",
	volume = "3",
	number = "08",
	pages = "S08002",
	year = "2008"
}

@article{spd,
    author = "Aamodt, K and others",
    authortype = {(ALICE Collaboration),},
    %title = "{Alignment of the ALICE Inner Tracking System with cosmic-ray tracks}",
    eprint = "1001.0502",
    archivePrefix = "arXiv",
    primaryClass = "physics.ins-det",
    doi = "10.1088/1748-0221/5/03/P03003",
    journal = "JINST",
    volume = "5",
    pages = "P03003",
    year = "2010"
}

@book{pmd,
      author        = "Cortese, P and others",
      authortype    = {(ALICE Collaboration),},
      %title         = "{ALICE Photon Multiplicity Detector (PMD): addendum to the Technical Design Report}",
      series        = "CERN-LHCC-2003-038",
      year          = "2003",
      %url           = "http://cds.cern.ch/record/642177",
}

@book{zdc,
    author        = "Gallio, M and Klempt, W and Leistam, L and De Groot, J and Schukraft, J",
    %title         = "{ALICE Zero Degree Calorimeter (ZDC): Technical Design Report}",
    authortype    = {(ALICE Collaboration),},
    year          = "1999",
    series        = "CERN-LHCC-99-005",
    %url           = "https://cds.cern.ch/record/381433"
}
